\documentclass[aps,pre,twocolumn,superscriptaddress,showpacs]{revtex4-1}
\usepackage{graphicx}
\usepackage{amsmath}
\usepackage{natbib}
\usepackage[usenames]{color}
\begin{document}
%--------------------------------------------------------------------
\title{Incommensurate standard map}
%--------------------------------------------------------------------

\author{Leonardo Ermann}
\affiliation{Departamento de F\'{\i}sica Te\'orica, GIyA,
 Comisi\'on Nacional de Energ\'{\i}a At\'omica.
 Av.~del Libertador 8250, 1429 Buenos Aires, Argentina}
\affiliation{Consejo Nacional de Investigaciones
Cient\'ificas y T\'ecnicas (CONICET), Buenos Aires, Argentina}

\author{Dima L. Shepelyansky}
\affiliation{Laboratoire de Physique Th\'eorique, IRSAMC, 
Universit\'e de Toulouse, CNRS, UPS, 31062 Toulouse, France}

%--------------------------------------------------------------------
\date{\today}
%\date{October ??, 2018}
%--------------------------------------------------------------------
\begin{abstract}
We introduce and study the extension of the Chirikov standard map
when the kick potential has two and three incommensurate spatial harmonics.
This system is called the incommensurate standard map.
At small kick amplitudes the dynamics is bounded
by the isolating Kolmogorov-Arnold-Moser surfaces while above a certain kick
strength it becomes unbounded and diffusive.
The quantum evolution 
at small quantum kick amplitudes is somewhat similar to the case of Aubru-Andr\'e
model studied in mathematics and experiments with cold atoms
in a static incommensurate potential.
We show that for the quantum map there is also
a metal-insulator transition
in space while in momentum
we have localization similar to the case of 2D Anderson localization.
In the case of three incommensurate frequencies of space potential the quantum evolution
is characterized by the Anderson transition similar to 3D case
of disordered potential.
We discuss possible physical systems with such map
description including dynamics of comets and dark matter in planetary systems.
\end{abstract}
%
%\pacs{05.45.Mt, 03.65.Sq}
%
\maketitle
%--------------------------------------------------------------------

%%%%%%%%%%%%%%%%%%%%%%
\section{Introduction}
\label{sec:1}
%%%%%%%%%%%%%%%%%%%%%%

The investigation of dynamical symplectic maps 
allows to understand the fundamental deep properties of 
Hamiltonian dynamics. The map description originates
from the construction of Poincar\'e sections
of continuous dynamics invented by Poincar\'e \cite{poincare}.
The mathematical foundations for symplectic maps
are described in \cite{arnold,sinai}. Their
physical properties and  applications
including numerical studies
are given in \cite{chirikov,lichtenberg}.
The renormalization features
of critical Kolmogorov-Arnold-Moser (KAM) 
invariant curves are analyzed in \cite{mackay}
with the transport properties
through destroyed KAM curves investigated in
\cite{chirikov,meiss}.

The seminal example of a symplectic map
is the Chirikov standard map \cite{chirikov}
\begin{eqnarray}
 \bar{p} = p + K \sin{x} \;, \;\;
 \bar{x} = x + \bar{p} \;,
\label{eq:stmap}
\end{eqnarray}
where $p$ and $x$ are canonically conjugated
variables of momentum $p$ and coordinate $x$
and bars mark the values of variables after a map iteration.
For $K > K_c=0.9716...$ the last KAM curve is destroyed and 
the dynamics is characterized by a global chaos
and diffusion in momentum $<p^2> = D t$
$p$ with a diffusion rate $ D $,
where the time $t$ is measured in number of map iterations
and the diffusion coefficient is $D \approx K^2/2$
at large $K$ values \cite{chirikov,mackay,meiss}.

The important feature of map (\ref{eq:stmap}) is its
universality related to an equidistant spacing between
resonance frequencies so that a variety
of symplectic maps and periodically driven Hamiltonian 
systems can be locally described by the map (\ref{eq:stmap})
in a certain domain of the phase space. The Chirikov standard map
finds applications for description of various
physical systems including plasma confinement in open mirror traps,
microwave ionization of hydrogen atoms,
comet and dark matter dynamics in the Solar System
and behavior of cold atoms in kicked optical lattices
(see e.g. 
\cite{chirikov,stmapscholar,raizenscholar,dlshydrogen,kepler} 
and Refs. therein).

The quantum version of map (\ref{eq:stmap}) 
is obtained by considering  $p$ and $x$ 
as the Heisenberg operators
with the  commutation relation $[\hat{p},\hat{x}]=-i \hbar$.
The corresponding quantum evolution of the wave function
$\psi (x)$ is described by the map \cite{cis1981,cis1988}:
\begin{eqnarray}
 \bar{\psi} = \exp(-i k \cos x) \exp(- i {\hat{p}}^2/2\hbar) \psi \; ,
\label{eq:qstmap}
\end{eqnarray}
where $\hat{p} = \hbar \hat{n} = - i \hbar \partial/\partial{x}$,
$k=K/\hbar$, $T=\hbar$, $K=kT$ ($T$ can be also considered as 
a rescaled time period between kicks). Here, the wave function is defined in 
the domain $0 \leq x \leq 2\pi$ corresponding to the quantum rotator case,
or it can be considered on the whole interval $-\infty <x < \infty$
corresponding to motion of cold atoms in an optical lattice.
The later case has been realized in experiments with 
cold atoms in kicked optical lattice \cite{raizen}.
At $K > K_c$ the classical diffusion in $p$
becomes localized by the quantum interference effects
with an exponential decay of probability over the momentum states $n = p/\hbar$:
\begin{eqnarray}
 <|\psi_n|^2> \propto \exp(-2 |n-n_0|/\ell) \; ; \; \ell \approx D/\hbar^2 
\label{eq:stloc}
\end{eqnarray}
with the localization length $\ell$ and $n_0$ being an initial state
\cite{cis1981,dls1987}. This dynamical quantum localization
is analogous to the Anderson localization in disordered
solids as pointed in \cite{fishman}. However,
the role of spacial coordinate is played by  momentum state level index $n$
and diffusion appears due to dynamical chaos in the classical limit
and not due to disorder 
(see more detail in \cite{fishmanscholar,cis1988,dls1987}).

The important feature of the Chirikov standard map
is its periodicity in spacial coordinate (of phase) $x$.
The cases with several kick harmonics
have been considered for various 
map extensions (see e.g. \cite{lichtenberg,chirikov,cis1981,typmap})
but all of them had periodicity in $x$. The important 
example of the map with several harmonics is the generalized
Kepler map which provides an approximate
description of the Halley map dynamics in the Solar System
\cite{halley} (see also \cite{kepler} for Refs. and 
extensions). In this system Jupiter gives an effective kick in energy
of the Halley comet when it passes through its perihelion.
This kick function contains several sine-harmonics
of Jupiter rotational phase
since the comet perihelion distance is inside the Jupiter orbit.
However, other planets, especially Saturn, also
give a kick change of comet energy. Since the frequencies of other
planets are generally not commensurate
with the Jupiter rotation frequency 
we have a situation where the kick function 
depends at least on two phases with incommensurate frequencies. 
Thus it is important to analyze the incommensurate extensions
of the Chirikov standard map.

The simplest model is the incommensurate standard (i-standard) map 
which we investigate in this work:
\begin{eqnarray}
\nonumber
 \bar{p}&=&p+K_1\sin{x}+K_2\sin{\nu x}\\
%\nonumber
 \bar{x}&=&x+\bar{p}
\label{eq:istand}
\end{eqnarray}
where $\nu$ is a generic irrational number,
and $K_1$, $K_2$ are amplitudes of two incommensurate
kick harmonics. 
Here the coordinate domain is $-\infty < x < \infty$
corresponding to dynamics of atoms in 
an incommensurate
optical lattice.
For $K_1=0$ or $K_2=0$ the model is reduced to the map (\ref{eq:stmap}).
We consider here the case of the golden mean value
$\nu=(\sqrt{5}-1)/2=0.618...$ . 

The incommensurate standard map (\ref{eq:istand})
describes the dynamics of cold atoms in a kicked optical lattice
with an incommensurate potential.
The static incommensurate optical lattice
can be created by laser beams with two incommensurate wave lengths.
In fact such incommensurate optical lattice had been already realized
in cold atoms experiments \cite{inguscio} where the evolution of atomic 
wavefunction can be  approximately described \cite{modugno}
by the Aubry-Andr\'e model on a discrete incommensurate 
integer lattice with the Aubry-Andr\'e transition from localized to delocalized
states \cite{aubryandre}. At present the investigation of
interactions between atoms on such a lattice attracts a significant
interest of cold atoms community (see e.g. \cite{bloch}).

Due to the above reasons we think that
the incommensurate standard map will capture
new features of dynamical chaos with possible
application to various physical systems.
We also note that the quantum evolution of
this map may have localization or delocalization properties 
with a certain similarity with the Anderson transition
in disordered solids.

The paper is constructed as follows:
Section II describes the properties of classical dynamics;
quantum map evolution is analyzed in Section III,
effective two-dimensional (2D) and three-dimensional
features  of quantum evolution are considered in Section IV,
the discussion of  results is given in Section V.

%%%%%%%%%%%%%%%%%%%%%%
\section{Classical map dynamics}
\label{sec:2}
%%%%%%%%%%%%%%%%%%%%%%

\begin{figure}
\includegraphics[width=0.475\textwidth]{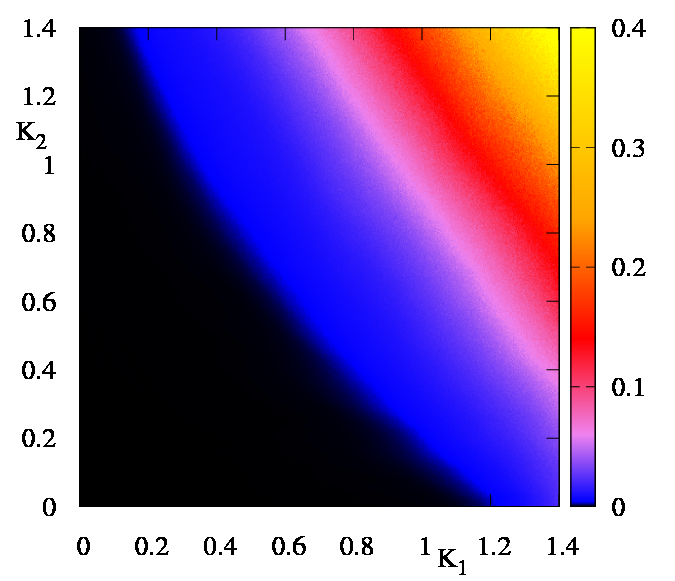}
\caption{Diffusion rate
in momentum 
$D=\langle p^2 \rangle/t$ vs $K_1$ and $K_2$.
Data are averaged over $1000$ trajectories with random initial 
conditions in the interval $x,p\in[0,10^{-6})$ for the number of iterations 
$t=10^4$. Color bar shows the values of $D$.
}
\label{fig1}
\end{figure}

To study the properties of classical dynamics of map (\ref{eq:istand})
we lunch a bunch of trajectories
in a vicinity of unstable fixed point
$x=0, p=0$ and compute an effective diffusion
coefficient $D=<p^2>/t$ averaged over all initial trajectories.
The dependence of $D$ on $K_1$ and $K_2$ is shown in Fig.~\ref{fig1}.
These data show that there is a critical curve
$K_{c2} = f(K_{c1})$ below which
the momentum oscillations ave bounded
and above which $p$ grows diffusely with time.
At $K_1=K_2=K$ we obtain $K_c = K_{c1}=K_{c2} \approx 0.65$ (see below).
Of course, the time $t$ used in Fig.~\ref{fig1}
is not very large (due to many trajectories and many $K_1, K_2$
values) so that we obtain only an approximate
position of the critical curve.
Thus for $K_2=0$ we know that $K_c=0.9716...$ \cite{mackay}
that is a bit below than the blue domain with
a finite diffusion rate $D$. This happens due to not very large
$t =10^4$ value and a small diffusion near $K_c$
being $D \approx 0.3 (K-K_c)^3$ for the map (\ref{eq:stmap}) \cite{chirikov,dls1987}.

\begin{figure}
\includegraphics[width=0.475\textwidth]{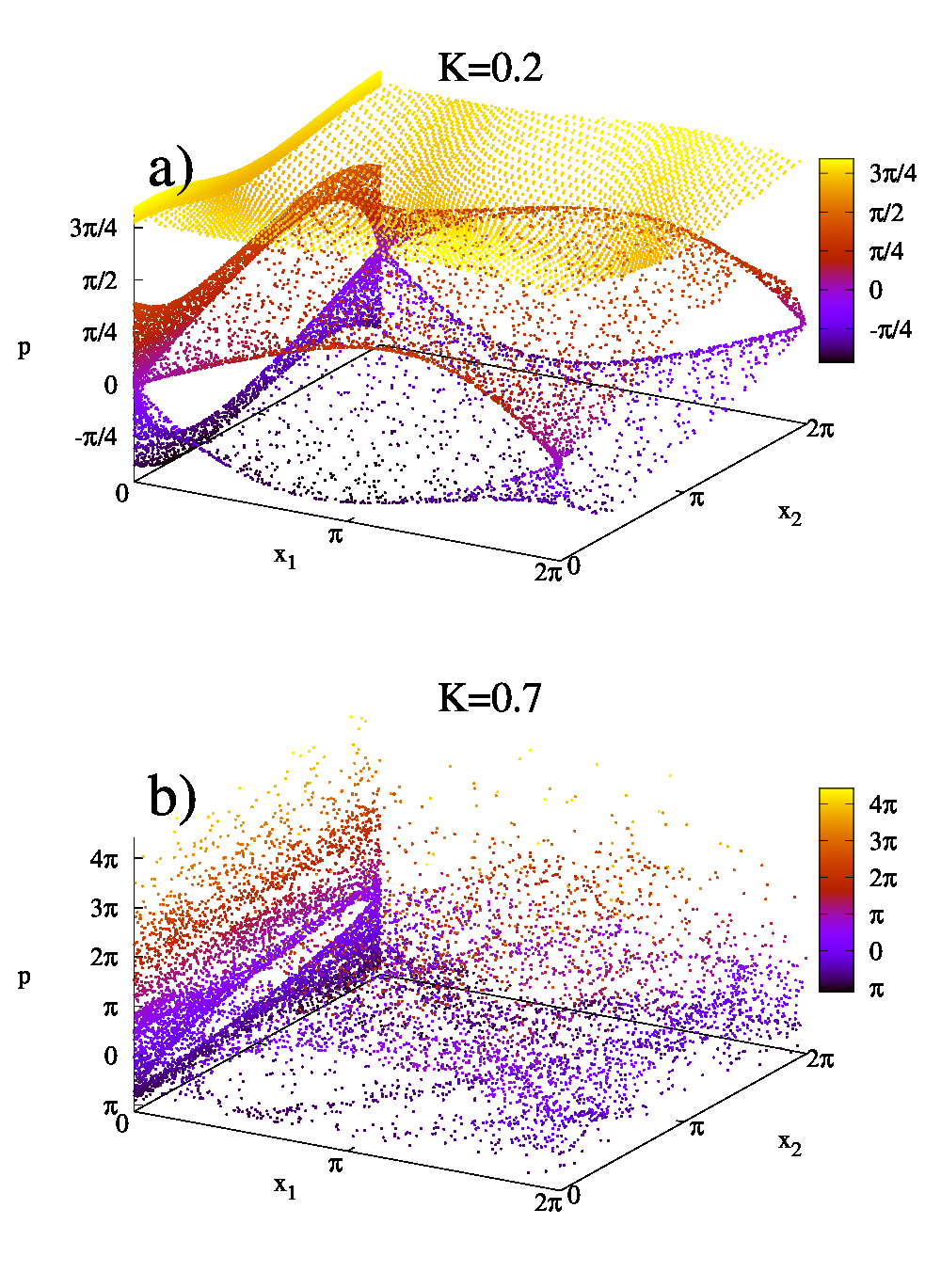}
\caption{Poincar\'e section in the phase space 
 $(x_1,x_2,p)$ for $K=K_1=K_2=0.2$ (a)
and $K=K_1=K_2=0.7$ (b).
a)The data are obtained from two trajectories
with $t=5000$ map iterations
for the initial values
$p=x_1=0.01, x_2 = 0.01 \times \nu$
with $p=0.01$ and $p=\pi/\sqrt{2}$;
b)the initial phases $x_1, x_2$ are the same as in (a)
and $p=0.01$.
Color bars show $p$ values.  
}
\label{fig2}
\end{figure}

To represent trajectories on the Poincar\'e section
it is convenient to use variables
$x_1 = x (\mod 2\pi)$, $x_2 = \alpha x (\mod 2\pi)$ and $p$.
We show two sections for $K = K_1 = K_2 $
  below the critical value at $K=0.2 < K_c$
and above the critical value at $K=0.7 > K_c$
(see Fig.~\ref{fig2}). For $K=0.2$ there are smooth
invariant KAM surfaces bounding $p$ variations
while for $K=0.7$ there is a chaotic unbounded dynamics in momentum.

\begin{figure}
\includegraphics[width=0.425\textwidth]{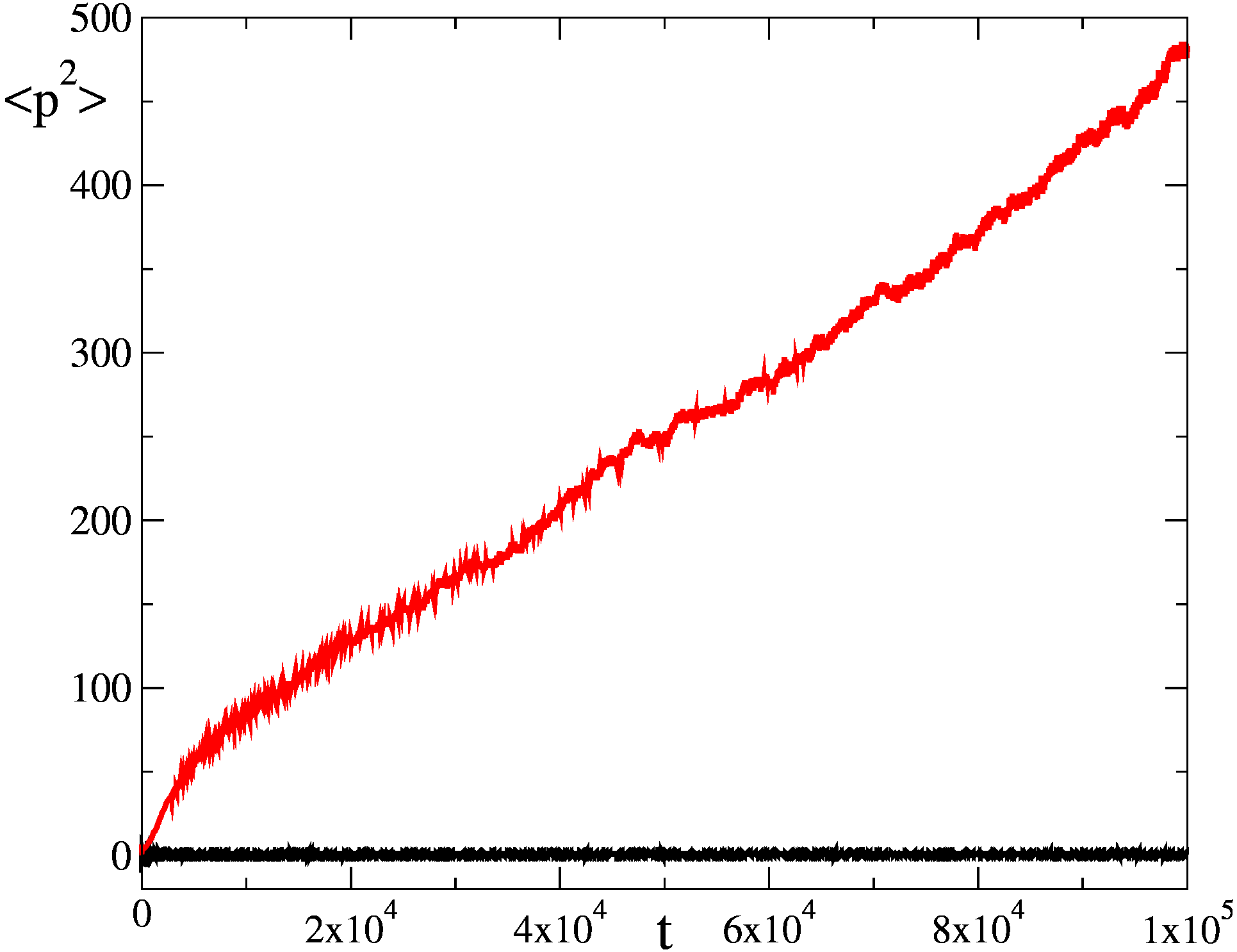}
\caption{Dependence of $\langle p^2 \rangle$ on time $t$  
for $K_1=K_2=0.2$ in black (bottom) and 
$K_1=K_2=0.7$ in red (top) solid curves.
The average is done over $1000$ trajectories with random initial
 conditions in the range $x,p\in[0,10^{-6})$ (similar to initial
conditions as in  Fig.~\ref{fig1}).
}
 \label{fig3}
\end{figure}

When the last KAM surface is destroyed we have a diffusive growth
of momentum as it is shown in Fig.~\ref{fig3} for $K=K_1 = K_2 =0.7 > K_c$.
For $K=K_1 = K_2 =0.2 < K_c$  the values of $p$ remain bounded for all times.

 \begin{figure}
\includegraphics[width=0.475\textwidth]{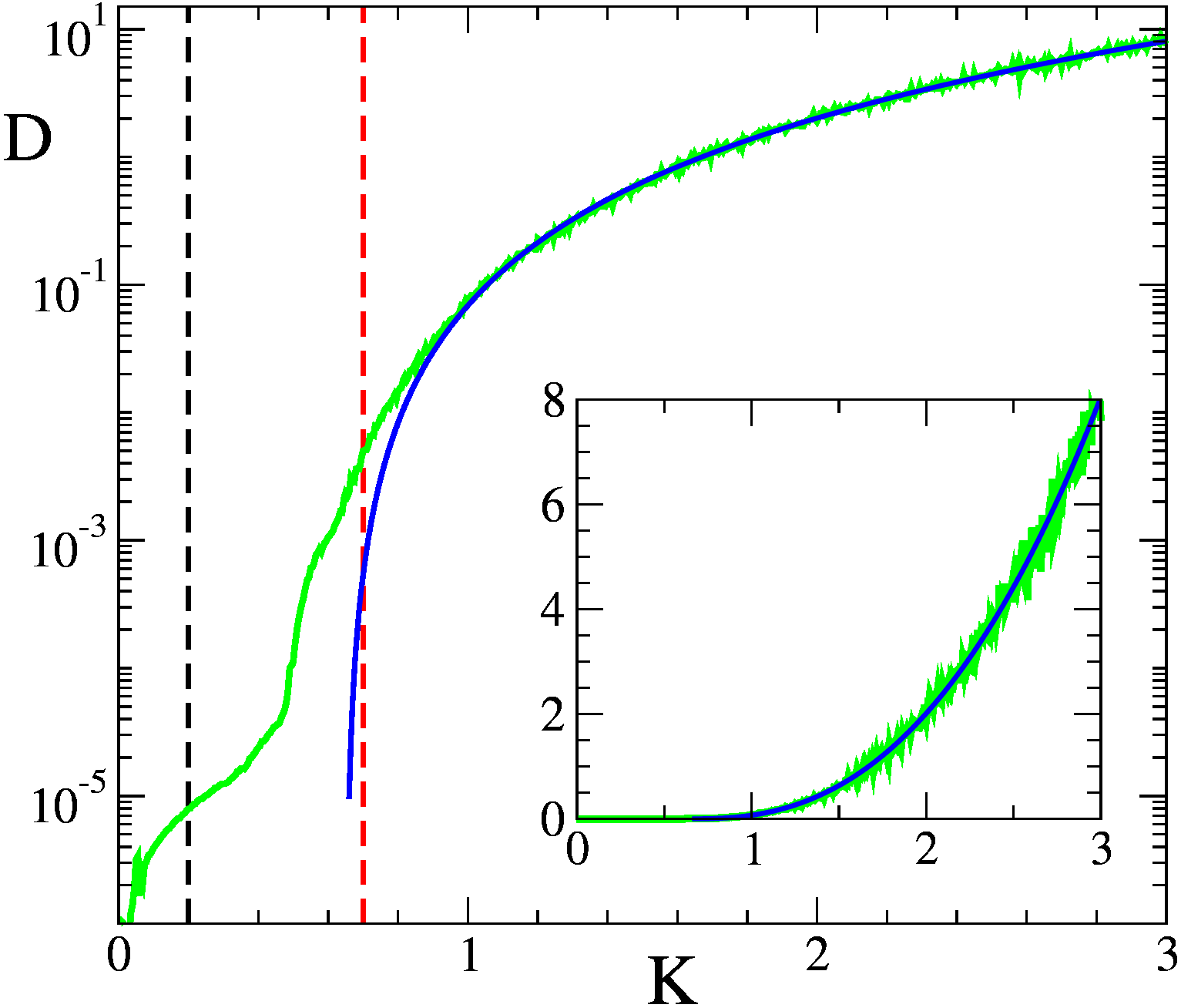}
\caption{Dependence of diffusion rate $D=\langle p\rangle/t$ on
$K=K_1=K_2$ shown by green (gray) curve, 
where vertical black and red (gray) dashed lines mark values 
$K=K_1=K_2=0.2$ and $K=K_1=K_2=0.7$ respectively. 
The fit dependence $D=D_0(K-K_c)^\alpha$ is shown by 
blue (black) curve with the fit values
$D_0=0.95\pm0.04$, $k_c=0.65\pm0.02$ and $\alpha=2.5\pm0.03$. 
Inset plot show the same curve for linear scale in $D$ axis.
Data are obtained from $1000$ trajectories and $t=10^5$ iterations.
}
 \label{fig4}
\end{figure}

From the time dependence of $\langle p^2 \rangle$ on time $t$
we compute the diffusion rate $D$. The dependence of $D$
on $K=K_1=K_2$ is shown in Fig.~\ref{fig4}.
We find that this dependence is satisfactory described by
the relation $D=D_0(K-K_c)^\alpha$
with $D_0 \approx 0.95$, $K_c \approx 0.65$ and
$\alpha \approx 2.5$. The value of the exponent $\alpha$
is close to the one for the Chirikov standard map
with $\alpha \approx 3$ \cite{chirikov,meiss,dls1987}.
Of course, the number of iterations $t=10^5$
is not very large and due to that we have only
approximate values of the fit parameters
$D_0, K_c, \alpha$. It is possible that 
the real value $K_c$ is a bit smaller then
the it value $K_c=0.65$ but more exact
determination of these values require 
separate studies.

With the obtained global properties of the
classical $i-$standard map we go to 
analysis of its quantum evolution in next Sections. 

%%%%%%%%%%%%%%%%%%%%%%
\section{Quantum map evolution}
\label{sec:3}
%%%%%%%%%%%%%%%%%%%%%%

The quantum evolution of $i$-standard map is described by 
the following transformation of wave function on one map period:
\begin{eqnarray}
\nonumber
 \bar{\psi} = \exp[-i (k_1 \cos x + (k_2/\nu) \cos \nu x)] \\
\times \exp[- i {\hat{p}}^2/2\hbar] \; \psi \; 
\label{eq:qistmap}
\end{eqnarray}
with $k_1=K_1/\hbar$, $k_2=K_2/\hbar$
and normalization condition ${\int^{\infty}}_{-\infty} |\psi(x)|^2 dx =1$. 
Here, the wave function $\psi$ evolution
is considered on infinite domain $-\infty < x <\infty$
corresponding to dynamics of cold atoms in kicked optical lattices;
the first multiplier describes kick from optical lattice
and the second one gives a free propagation in empty space.
The kick potential is $V(x)=k_1 \cos x + (k_2/\nu) \cos \nu x$.

\begin{figure}
\includegraphics[width=0.49\textwidth]{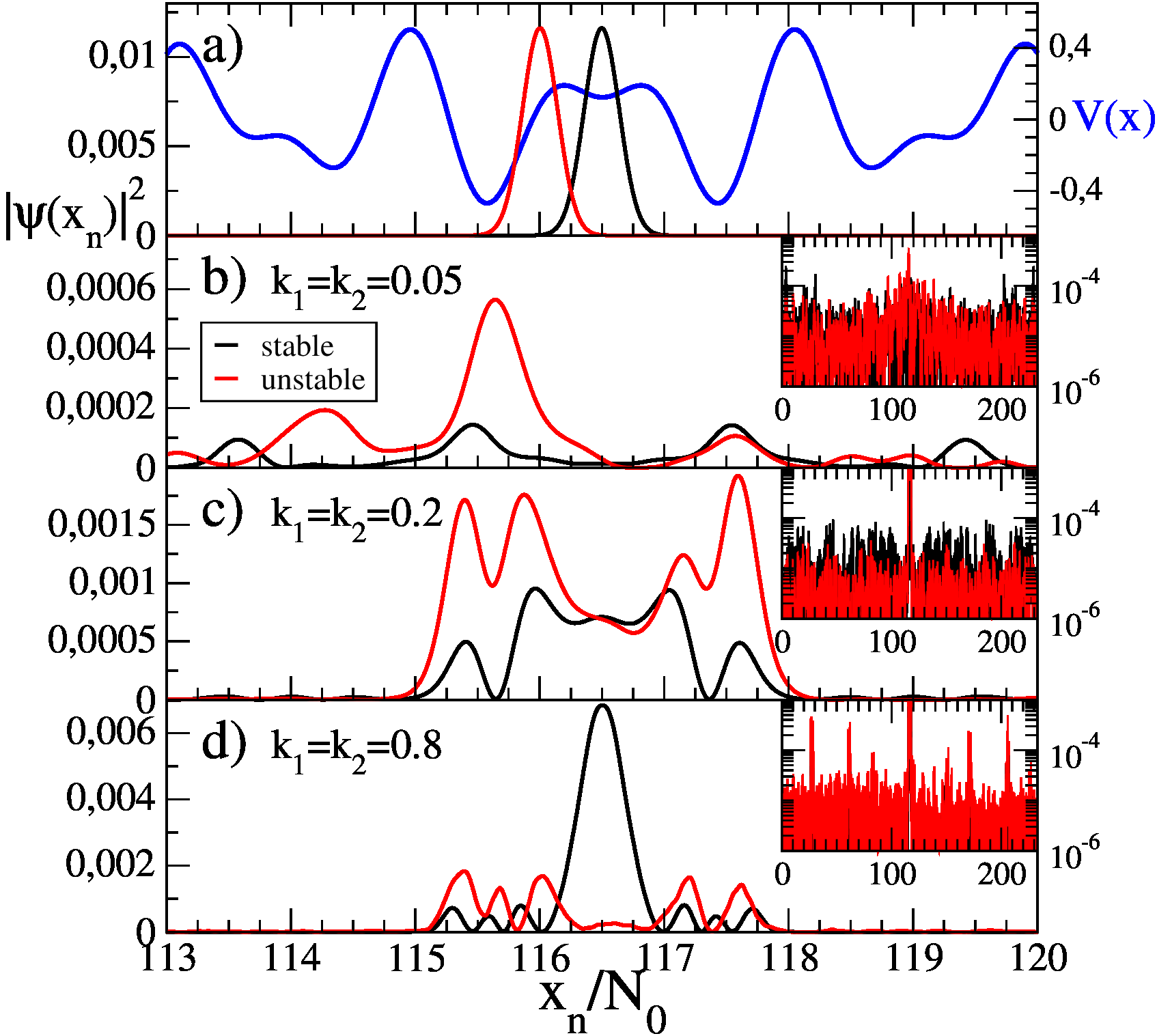}
\caption{Probability distribution in coordinate. 
Panel (a) shows the probability distribution in coordinate 
for initial state given by a Gaussian centered 
at stable (black curve) and unstable (red/gray curve) conditions. 
Blue (dark black) curves shows the potential energy $V(x)$ for 
$k_1=k_2=0.2$ where the scale is shown in right side.
Panels (b), (c), (d) (from top to bottom) show respectively the distributions for 
$k_1=k_2=0.05$ $k_1=k_2=0.2$ and $k_1=k_2=0.8$ after $t=10^4$ 
iterations with black and red curves showing the cases of stable and unstable 
initial conditions. 
In each panel, top-right insets
show the same distribution in logarithmic scale and for the whole Hilbert space.
In all cases $T=76\pi/3^5 \approx 0.983$.
}
 \label{fig5}
\end{figure}

To perform numerical simulations of the quantum map (\ref{eq:qistmap})
we approximate $\nu$ by its Fibonacci series with 
$\nu_m=r_m/q_m = 144/233$ considering time evolution
on a ring of size $[0, 2\pi q_m)$
with periodic boundary conditions. Here $r_m=f_{m-1}$ and $q_m=f_{m}$, 
are the $(m-1)^{\text{th}}$ and $m^{\text{th}}$ Fibbonaci numbers given 
by $f_m=f_{m-1}+f_{m-2}$ 
with $f_1=f_2=1$, and therefore $\lim_{m\rightarrow\infty}\nu_m=\nu$.
Then one iteration of the map is obtained as 
\begin{equation}
\bar{\psi}={\hat F^{-1}}e^{-i\left[k_1\cos{\left(\frac{2\pi x_n}{N_0}\right)}+
\frac{k_2}{\nu_m}\cos{\left(\frac{2\pi \nu_m x_n}{N_0}\right)}\right]} 
{\hat F}e^{-i \frac{T}{2}(\frac{p_n}{q_m})^2}\psi
\label{eq:qmapdiscrete}
\end{equation}
where $T=\hbar$ and the Hilbert space dimension is $N=N_0\times q_m$, 
with $q_m$ ($r_m$)  periodic space cells of internal 
dimension $N_0$ for $k_1$ ($k_2$) harmonic of $V(x)$.
Here positions and momentum have integer values $x_n=n$ and 
$p_n=(n-\lfloor N/2\rfloor)$
with $n=0,\ldots,N-1$; and ${\hat F}$ and ${\hat F^{-1}}$ are the operators of 
discrete Fourier transform
from momentum to coordinate representation and back.
For numerical simulations we have chosen  
$N_0=3^5=243$, $r_m=144$ and $q_m=233$, with $\hbar=\frac{76\pi}{3^5}\simeq 0.983$.

 \begin{figure}
\includegraphics[width=0.49\textwidth]{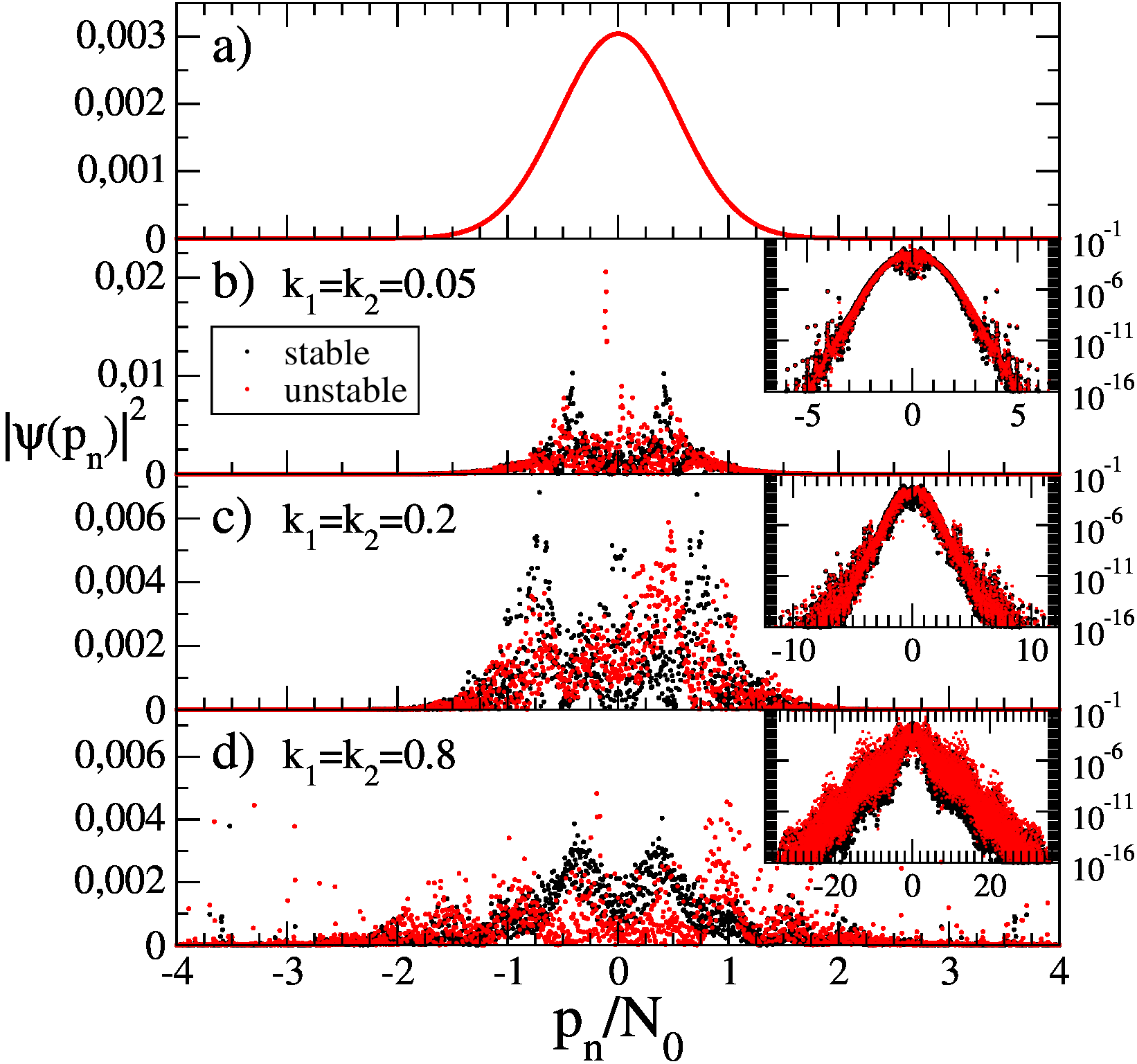}
\caption{Probability distribution 
in momentum for the same cases than \ref{fig5}.
Top panel show the probability distribution in momentum 
for initial state given by a Gaussian centered 
at zero (the same for stable and unstable conditions). 
In second, third and fourth panels (from top to bottom) 
the distribution is shown for $k_1=k_2=0.05$, $k_1=k_2=0.2$ 
and $k_1=k_2=0.8$ after $t=10^4$ iterations
where black and red dots illustrate the cases of stable and unstable conditions. 
In each panel, top-right insets
show the same distribution in logarithmic scale and for the whole Hilbert space. 
In all cases $T=76\pi/3^5 \approx 0.983$.
}
 \label{fig6}
\end{figure}

We choose as  an initial configuration a Gaussian wave packet 
centered at stable (or unstable) point of the kick potential
$V(x)= k_1 \cos{ x }+ (k_2/\nu) \cos{ \nu x}$. 
On a  discrete lattice of $x_n=n$ ($0 \leq n <N$)
this distribution is 
$  \psi(x_n) = A \exp[-(x_n-X_0)^2/(2 (N_0/5))^2)]$ with $\langle x \rangle=X_0$ 
and
$A$ a normalization factor.
The corresponding distribution, in momentum
space $p_n$, is obtained by the discrete Fourier transform, 
where in this case $\langle p \rangle=P_0=0$.

We define $P_{3,x}(t)$ and $P_{3,p}(t)$ as 
the probability to stay in 3 cells centered at initial 
$X_0$ and $P_0$ values respectively:
\begin{eqnarray}
 P_{3,x}(t)&=&\sum_{n=\left\lfloor X_0-\frac{3N_0}{2}\right\rfloor}
^{\left\lfloor X_0+\frac{3N_0}{2}\right\rfloor}\vert\psi(x_n)\vert^2
\label{eq:x0state}
\end{eqnarray}
\begin{eqnarray}
 P_{3,p}(t)&=&\sum_{n=\left\lfloor P_0-\frac{3N_0}{2}\right\rfloor}
^{\left\lfloor P_0 +\frac{3N_0}{2}\right\rfloor}\vert\psi(p_n)\vert^2\
\label{eq:p0state}
\end{eqnarray}
where $\lfloor x \rfloor$ is the integer part of x.

The initial probability distributions
placed in a vicinity of stable and unstable fix points
of the kick potential $V(x)$ are shown in Fig.~\ref{fig5}.
We note that for small $k_1 \sim k_2 \ll 1$
the map approximately describes a continuous
time evolution in a static potential.
In \cite{modugno} it is shown that in this case the quantum evolution
is approximately reduced to the Aubry-Andr\'e model on
a discrete lattice with the eigenstates of
the  stationary Schr\"odinger equation:
\begin{equation}
\begin{array}{c}
\lambda \cos(\hbar n+\beta)\phi_{n} + \phi_{n+1}+\phi_{n-1}=E\phi_{n} \; .
\end{array}
\label{eq:aubry}
\end{equation}
Here $\lambda$ is an effective dimensional energy 
of the  quasiperiodic potential and the hopping amplitude
being unity. A metal-insulator transition (MIT) takes place
from localized states at $\lambda>2$ to
delocalized eigenstates at $\lambda<2$ \cite{aubryandre}.
A review of the properties of the Aubry-Andr\'e model can be found 
in \cite{sokoloff} and the mathematical prove of  MIT is given in 
\cite{lana1}. An estimate obtained in  \cite{modugno}
shows that $\lambda \propto k_2 \exp(-C_1/{k_1}^{C_2}+2  C_3 \sqrt{k_1})$ 
for an irrational $\nu \sim 1$ with $C_1, C_2, C_3$ being numerical constants
order of unity. This approximate reduction of
the Schr\"odinger equation in a continuous quasiperiodic
potential $V(x)$ to the discrete lattice Aubry-Andr\'e model 
have been used in experiments with cold atoms
where the MIT was found at $\lambda=2$ \cite{inguscio,bloch}.

\begin{figure}
\includegraphics[width=0.49\textwidth]{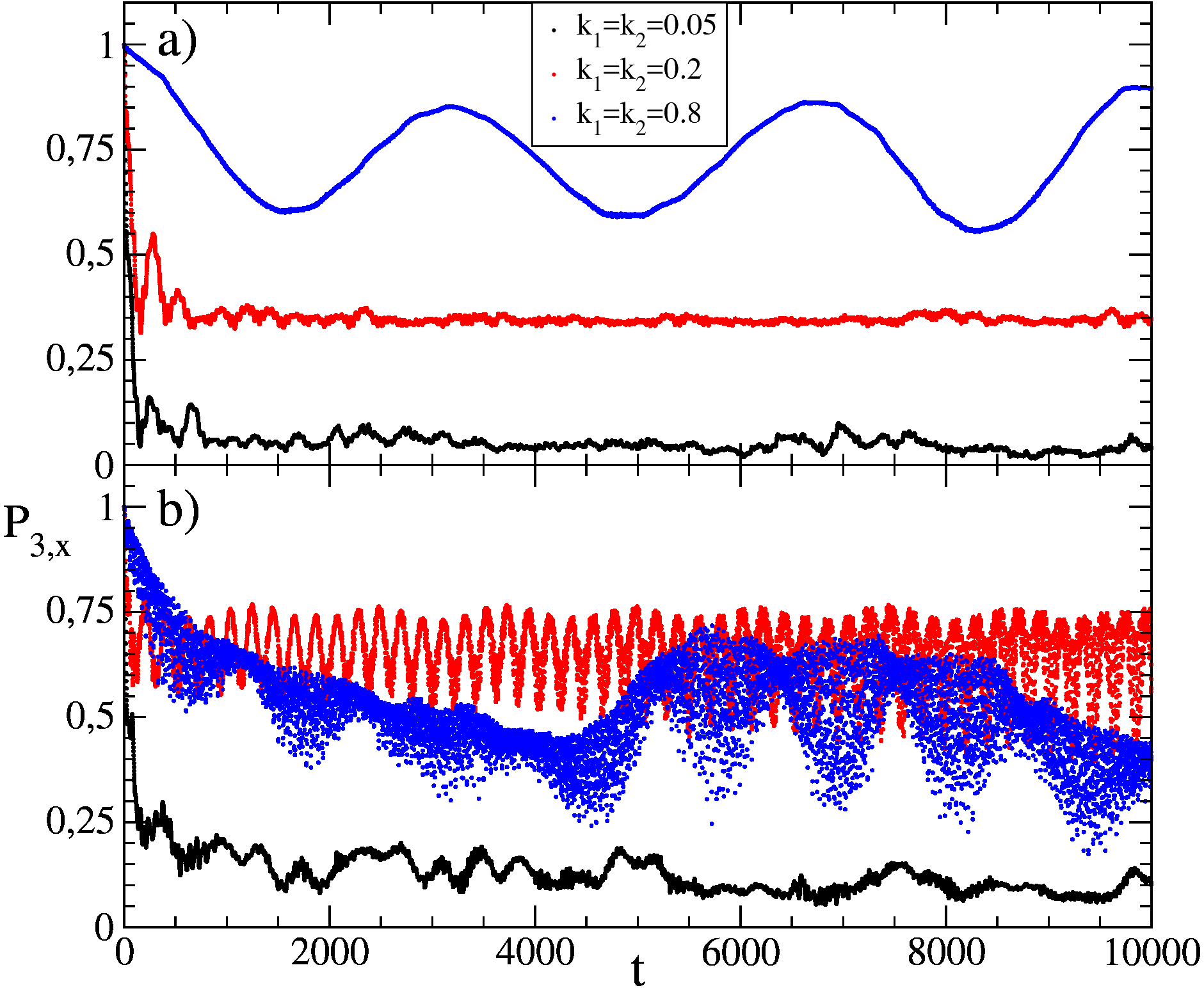}
\caption{Time evolution of the probability to stay 
in the first 3 space cells   $P_{3,x}$ 
for the cases of Figs \ref{fig5} and \ref{fig6} with $N_0=243$ and $q_m=233$.
Top (a) and bottom (b) panels show the stable and unstable conditions
of Fig.~\ref{fig5} 
respectively for $k_1=k_2=0.05$ (black circles), 
$k_1=k_2=0.2$ (red circles) and $k_1=k_2=0.8$ (blue circles)
ordered from bottom to top in each panel.
}
 \label{fig7}
\end{figure}

The signs of MIT are visible in Fig.~\ref{fig5}
with a delocalization of probability 
in space at $k_1=k_2=0.05$
and localization at $k_1=k_2=0.2; 0.8$.
At the same time the results of Fig.~\ref{fig6} show
that the probability distribution in momentum 
remains exponentially localized for the above $k_1,k_2$ values.

\begin{figure}
\includegraphics[width=0.49\textwidth]{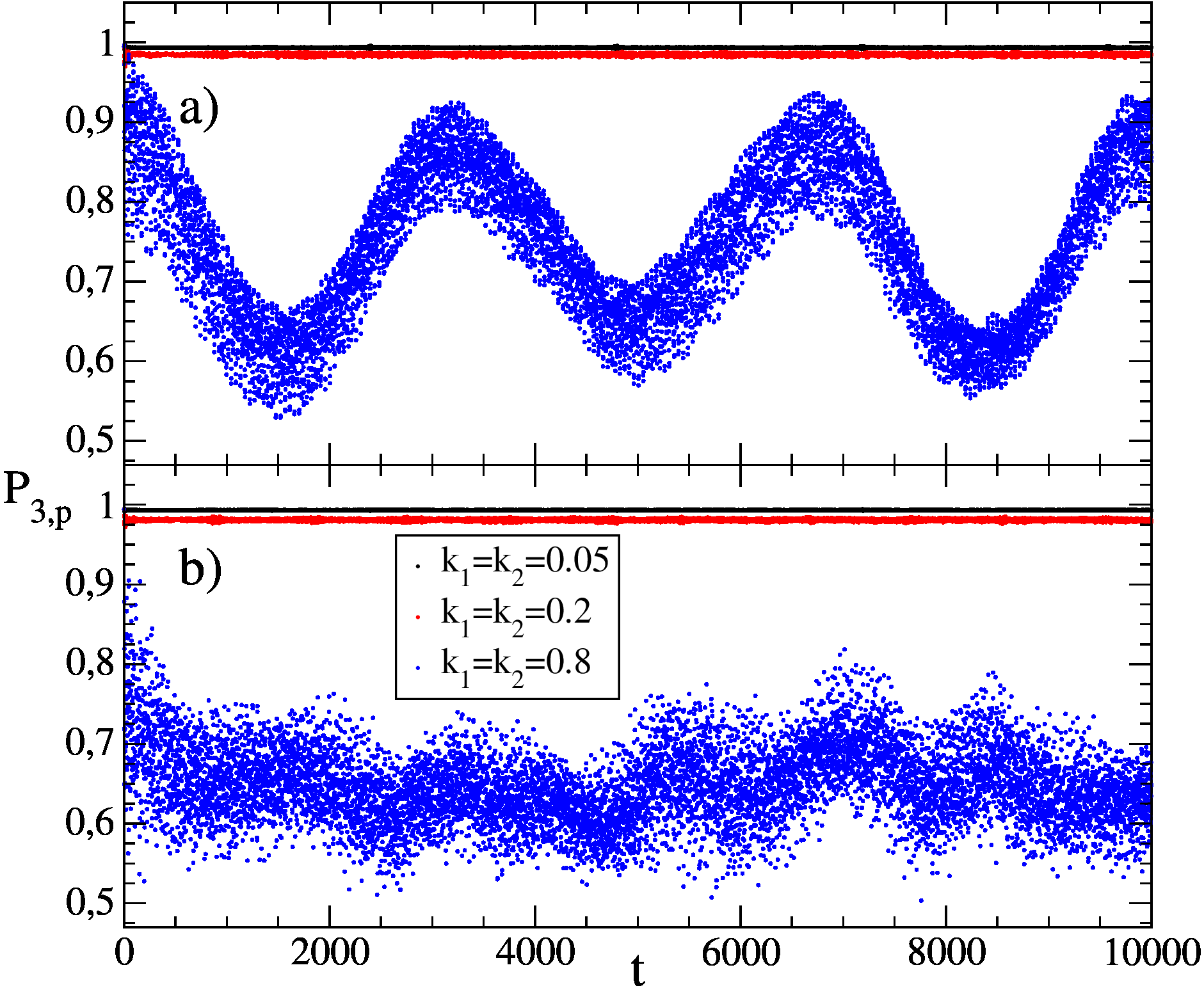}
\caption{Time evolution of the probability 
to stay in the first 3 momentum cells  $P_{3,p}$ 
for the cases of Figs \ref{fig5} and \ref{fig6} with $N_0=243$ and $q_m=233$.
Top (a) and bottom (b) panels show the stable and unstable conditions 
of Fig.~\ref{fig6}
respectively for $k_1=k_2=0.05$ (black circles), 
$k_1=k_2=0.2$ (red circles) and $k_1=k_2=0.8$ (blue circles)
ordered from top to bottom in each panel.
}
 \label{fig8}
\end{figure}

The time evolution of probabilities of stay
in a vicinity of initial cell 
$P_{3,x}$ and $P_{3,p}$ are shown in 
Fig.~\ref{fig7} and Fig.~\ref{fig8}.
We see that at $k_1=k_2=0.05$ 
only a small fraction of probability (about 10 percent)
remains
in a vicinity of initial space cell 
while in contrast it is rather large for 
 $k_1=k_2=0.2; 0.8$. In contrast
for the distribution in momentum 
about 99 percent 
remains in a vicinity of initial cell for
$k_1=k_2=0.05; 0.2$ and 65 percent for
$k_1=k_2=0.8$. Thus this data
confirms localization
in momentum and MIT
transition is space similar to
the  MIT discussed in \cite{modugno}
in the limit of energy conservative system
corresponding to the dynamics of 
our kicked model at small $k_1,k_2$ values.

\begin{figure}
\includegraphics[width=0.48\textwidth]{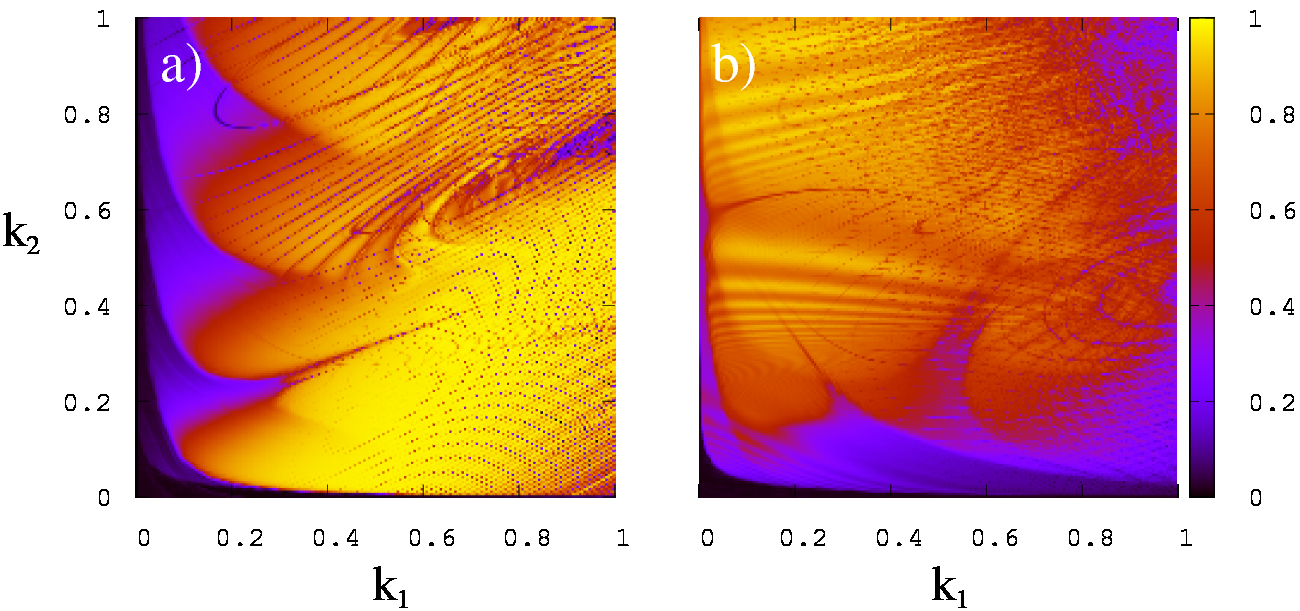}
\caption{ Probability $P_{3,x}(t)$ averaged in time within interval
$t\in (9000,10000]$ is shown as a function of $k_1$ and $k_2$.
Here $T=0.983...$ and the initial condition is centered at stable 
and unstable points of the potential in (a) and (b) panels respectively. 
}
 \label{fig9}
\end{figure}

\begin{figure}
\includegraphics[width=0.48\textwidth]{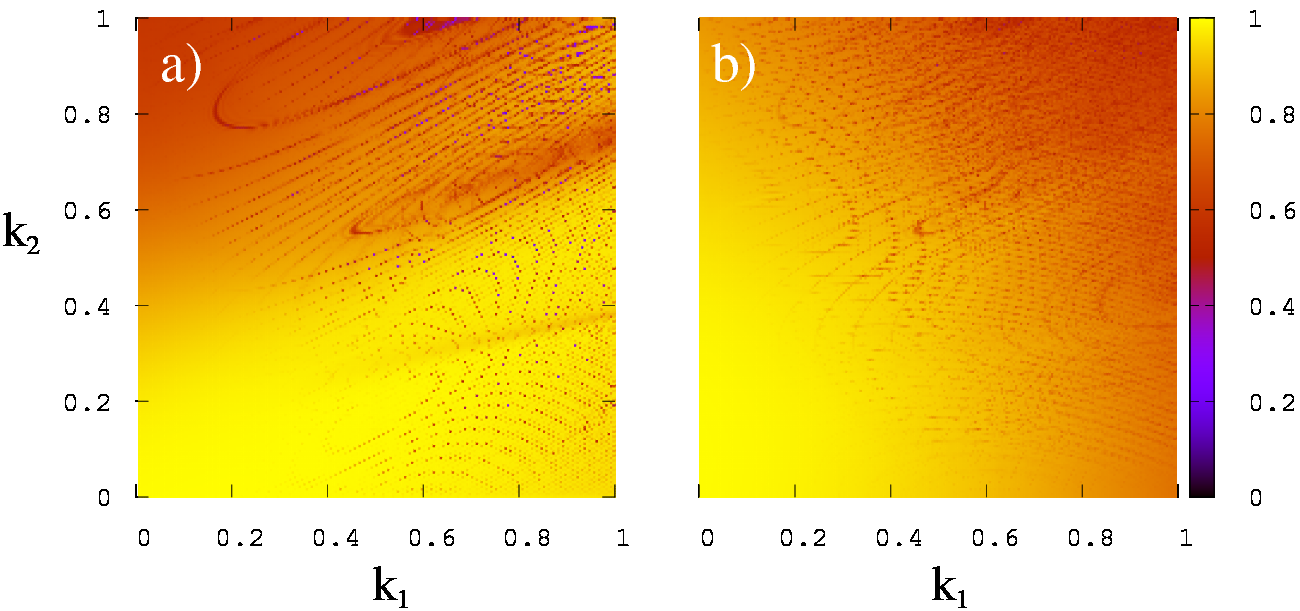}
\caption{Probability $P_{3,p}(t)$ in momentum averaged in time within interval
$t\in (9000,10000]$ is shown as a function of $k_1$ and $k_2$.
Here $T=0.983...$ and the initial condition is centered at stable 
and unstable points of the potential in left (a) and right (b) 
panels respectively. 
}
 \label{fig10}
\end{figure}

The global dependence of probabilities $P_{3,x}(t)$ and $P_{3,p}(t)$
on kick amplitudes $k_1,k_2$ are shown in Figs.~\ref{fig9},\ref{fig10}
for initial packet centered in a vicinity
of stable or unstable point. For $P_{3,x}(t)$
there is a clear region on $k_1,k_2$-plane where 
the probability $P_{3,x}(t)$ drops significantly
corresponding to delocalization in space (see Fig.~\ref{fig9})).
In contrast, the probability $P_{3,p}(t)$ remains always 
rather high showing that the classical 
chaotic diffusion in momentum is localized by quantum interference effects.

\begin{figure}
\includegraphics[width=0.48\textwidth]{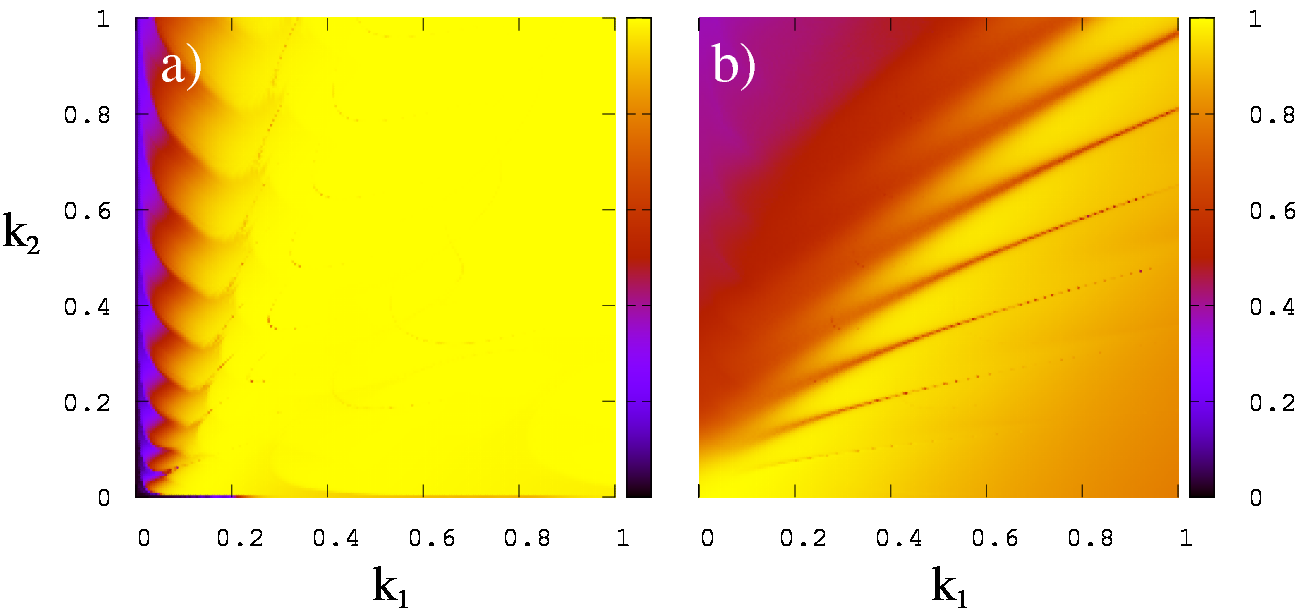}
\caption{Probability $P_{3,x}(t)$ in space (a) and $P_{3,p}(t)$ in momentum 
(b) averaged in time within interval
$t\in (9000,10000]$ are shown as a function of $k_1$ and $k_2$.
The initial condition in $x$ is centered at \it{stable} 
points of the potential and $T=0.2$. 
}
 \label{fig11}
\end{figure}

In Fig.~\ref{fig11} we consider a case with a smaller value
of $T=0.2$ so that the system becomes more close to the
case of stationary potential analyzed in \cite{modugno}.
This data are similar to the case at larger $T=0.983...$
with a domain of small probability  $P_{3,x}(t)$ values at small
$k_1,k_2$ indicated the delocalization of probability in space.
At the same time the probability $P_{3,p}(t)$ in momentum 
remains always localized.

The obtained results show the presence of 
certain space delocalization of the quantum
incommensurate map at small kick amplitudes
while at large amplitudes the probability remains
localized. The probability in momentum
remains localized for all $k_1,k_2$ values
at irrational $T/2\pi$ values.

At the same time we note that the rigorous
prove of probability delocalization in 
our model remains a mathematical challenge
since the map of our system
(or even its stationary version considered in \cite{modugno})
on the Aubry-Andr\'e model (\ref{eq:aubry})
works only approximately. Indeed, in our
model it is possible to have excitation of 
high energy states (even if the numerical results indicate
localization in momentum) that complicates the dynamics.
We hope that the skillful mathematical tools developed in 
\cite{lana1} will allow to obtain mathematical
results for the quantum incommensurate standard map.

Below we consider the localization properties 
in momentum in more detail.

\section{2D and 3D models of {\it i-}standard map}
 \label{sec:4}

To analyze the localization properties in momentum
we note that the kick with $k_1$ generates 
integer harmonics of $\exp(-i j_i x)$
while the kick with $k_2$ generates only harmonics
$\exp(-i j_2 \nu x)$ with integer $j_1, j_2$ values.
Due to that the wave function contains only these
two types of harmonics and the system
evolution is described by 
\begin{eqnarray}
\bar{\psi} = {\hat F^{-1}}  e^{-i [k_1\cos{x}+k_2\cos{\nu x}]}
{\hat F} e^{-i T(j_1+ \nu j_2)^2/2} \psi \; ,
\label{eq:kr2d}
\end{eqnarray}
where ${\hat F}$ is the 2D-fast Fourier transformation
from momentum to space representation
and ${\hat F^{-1}}$ gives a back transformation
from space to momentum. The integers $j_1$ and $j_2$
number the correspondent harmonic numbers
with the energy phase of free propagation between kicks
being $\phi_E = Tp^2/2=T(j_1+ \nu j_2)^2/2$
with $p=(j_1+\nu j_2)$.
If we would have $\phi_E(j_1,j_2)$ taking random values for each $j_1,j_2$
then we would have 2D kicked rotator with the
Anderson type localization in 2D.
In such a case we would expect that
the localization length $\lambda$ grows 
exponentially with the diffusion rate
$\ln \lambda \sim D \sim ({k_1}^2 + {k_2}^2)/2$
(see e.g. \cite{dls1987,borgonovi,imry}).
However, the phases  $\phi_E(j_1,j_2)$ are not random
but incommensurate and the appearance
of 2D Anderson
localization is not so obvious.

\begin{figure}
\includegraphics[width=0.475\textwidth]{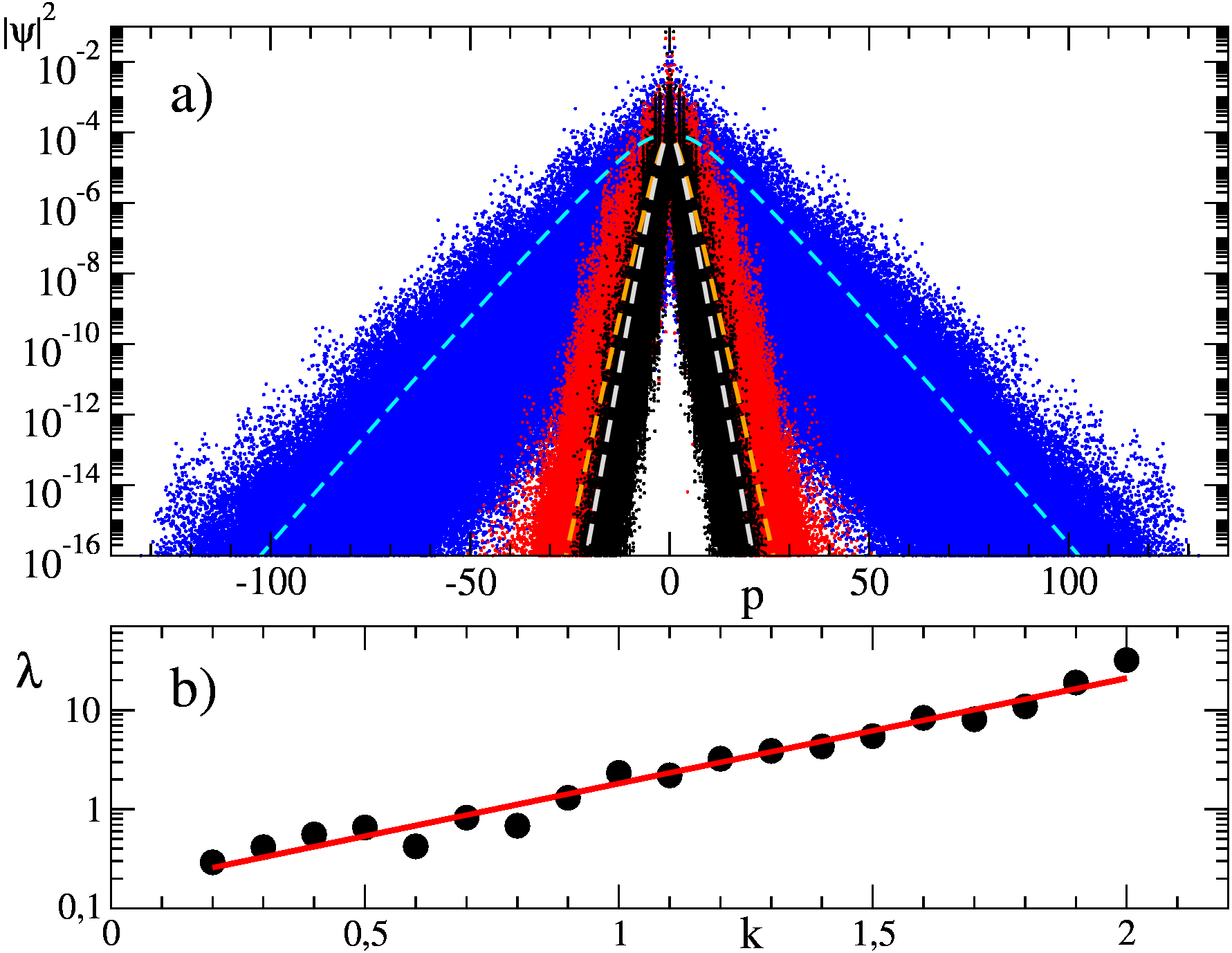}
\caption{(a) Probability distribution $\vert\psi(j_1,j_2)\vert^2$ 
in the  model (\ref{eq:kr2d}) as a function of $p=j_1+\nu j_2$
shown at time $t=10^4$.
Here $T=2$ and $N_2=3^7=2187$ ($N=4782969$).
Black (internal domain), 
red/gray (middle domain) and 
blue/dark (external domain) circles represent $k=k_1=k_2=0.5,0.7,1.2$ respectively. 
Numerical fits of $\vert\psi \vert^2\sim e^{-\frac{p}{\lambda}}$
with 
localization length values are shown by dashes lines
with $\lambda{(k=0.5)} \approx 0.656$, 
$\lambda{(k=0.7)} \approx 0.822$ and $\lambda{(k=1.2)} \approx 3.22$. 
Panel (b) shows the exponential dependence of fitted values of $\lambda$ as a function 
of $k$. Numerical fit is shown
by the straight line with the exponential growth $\lambda \approx 0.16 \exp{(2.44 k)}$.
Initial state is at $j_1=j_2=0$.
}
\label{fig12}
\end{figure}

For investigation of this expected 2D localization we take $N_2=3^7$
harmonics $j_1$ and $j_2$ so that the total number of states
becomes $N={N_2}^2=4782969$. The results of numerical show an exponential 
decay of probability with momentum $p$
as it is shown in Fig.~\ref{fig12}(a) for a few $k=k_1=k_2$ values.
We fit this decay by an exponential dependence $|\psi(p)|^2 \sim \exp(-|p|/\lambda)$
thus determining the localization length $\lambda$.
The results presented in Fig.~\ref{fig12}(b).
show the expected exponential growth of localization length
$\ln \lambda \sim 2.3 k$. The fact that $\ln \lambda $ is proportional
to $k$ and not to expected $k^2$ can be attributed to the fact
that we are still relatively close to the chaos border
(see Fig.~\ref{fig4}) and that the diffusion rate is small
while the estimate $\ln \lambda \sim D \sim k^2$ assumes 
well developed chaotic regime with a relatively
high $D$ \cite{imry}.

\begin{figure}
\includegraphics[width=0.475\textwidth]{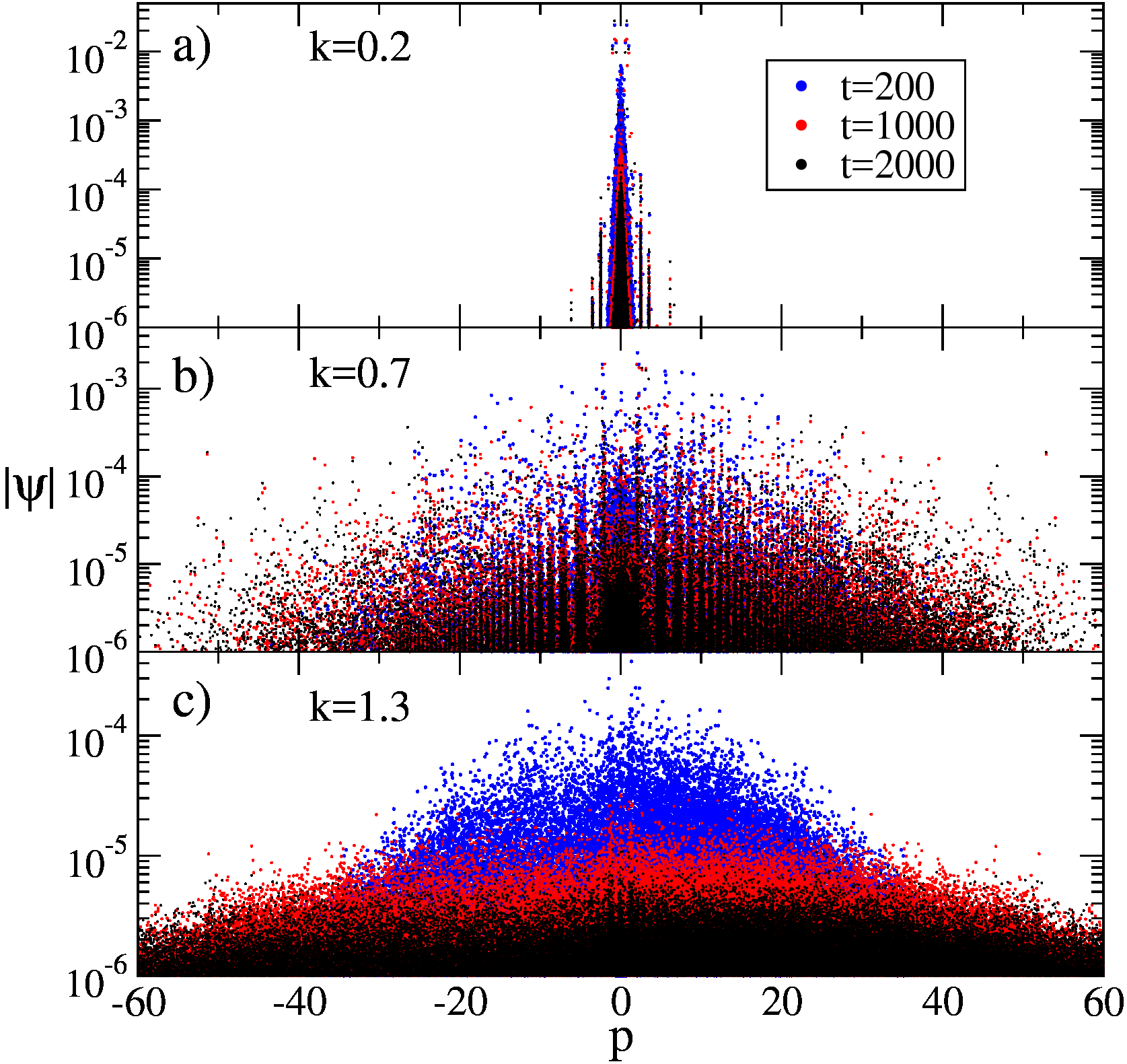}
\caption{Probability distribution $\vert\psi(j_1,j_2,j_3)\vert^2$ 
in the model (\ref{eq:kr3d}) as a function of $p=j_1+ j_2/\theta+ j_3/\theta^2$
where $\theta$ is the solution of $\theta^3-\theta-1=0$. Here
$T=2$ and $N_3=529$ ($N={N_3}^3=148035889$) at  evolution times 
$t=200, 1000, 000$ with blue (top dark points), 
red (middle gray points) and black (bottom) points respectively. 
Top (a), middle (b) and bottom (c) panels show the cases of 
$k_1=k_2=k_3=0.2$, $k_1=k_2=k_3=0.7$ and $k_1=k_2=k_3=1.3$.
Initial state is at $j_1=j_2=j_3=0$. The Anderson transition takes place
at $k = k_1 =k_2 \approx 0.7$.
}
 \label{fig13}
\end{figure}

There is no delocalization in 2D 
but in 3D there is the Anderson transition 
to delocalization \cite{anderson}
if a disorder is below a critical value 
or chaotic diffusion rate is higher a certain border 
(see e.g. \cite{imry}).
We argue that 3D case can be realized in 
our model if we add kick with one more 
incommensurate potential
$V_3(x)=k_3 \cos \nu_3 x$.
Then the wave function  additional harmonics
$\exp(-i j_3 \nu_3 x)$ and in analogy with (\ref{eq:kr2d})
the time evolution is described by
\begin{eqnarray}
\nonumber
\bar{\psi}&=&
 {\hat F^{-1}}  e^{-i [k_1\cos{x}+k_2\cos{\nu x}+k_3\cos{\nu_3 x]}}\\ 
&\times& {\hat F} e^{-i T(j_1+ \nu j_2 + \nu_3 j_3)^2/2} \psi \; ,
\label{eq:kr3d}
\end{eqnarray}
where momentum integer harmonics are  $j_1,j_2,j_3=1,...,N_3$ and
with the total dimension $N=N_3^3=529^3 = 148035889$ and 
$p=j_1+j_2/\theta+ j_3/\theta^2$, with
irrational $\theta = 1.32471795724475...$ 
being the solution of equation $\theta^3-\theta-1=0$.
Thus $\nu =1/\theta$, $\nu_3=1/\theta^2$.

The results for time evolution of probability
are presented in Fig.~\ref{fig13}.
They clearly show that at $k=k_1=k_2=0.2$
there is exponential Anderson localization of probability over momentum 
\cite{anderson}. For $k=k_1=k_2=1.3$ there is
spreading of probability in time over momentum states.
The case at  $k=k_1=k_2=0.7$ is close to a critical parameter
value where the Anderson transition takes place.
Thus the MIT point is located in the range
$0.7 \leq k_c <1.3$. Additional studies should be performed
to obtain the critical parameter more exactly
but the presented results definitely show that the transition takes 
place for $k$-parameter in this range.
 
 %%%%%%%%%%%%%%%%%%%%%%
\section{Discussion}
\label{sec:5}
%%%%%%%%%%%%%%%%%%%%%%

In our research we determined the main
properties of the incommensurate standard map
for its classical dynamics and for its
quantum evolution.
In the classical case the invariant KAM surfaces are destroyed 
above certain kick amplitudes given us a critical
curve on the plane of kick amplitudes $K_1,K_2$ (see Fig.~\ref{fig1}).
We find that above the critical curve
at its vicinity the diffusion rate
is characterized by a critical exponent $\alpha \approx 2.5$
which is not so far from the case
of the Chirikov standard map.

The quantum evolution at small 
quantum kick amplitudes $k_1=K_1/\hbar$, $k_2=K_2/\hbar$
is similar to the Aubry-Andr\'e type transition \cite{aubryandre}
as discussed in \cite{modugno} and observed in
cold atom experiments with a static incommensurate 
potential \cite{inguscio,bloch}.
However, at larger values of $k_1, k_2$ the evolution
remain localized  both in space and momentum.
We show that the localization in momentum
is similar to the case of Anderson localization in 2D.
While a significant progress has been reached with
rigorous results for Aubry-Andr\'e model \cite{lana1},
we point that the mathematical prove of 
space and momentum localization
for the quantum incommensurate standard map
represents a high challenge for mathematicians.

The quantum evolution
for the quantum $i-$standard map is
always localized in momentum,
as in the case of 2D Anderson localization.
However, for three kick-harmonics the situation
becomes similar to the 3D Anderson localization
with MIT from localized to delocalized 
evolution as the kick amplitude is increased.
We note that this behavior has similarities with
the frequency modulated kicked rotator
introduced in \cite{dls1983}
which also demonstrates the Anderson localization
in effective 2 and 3 dimensions \cite{borgonovi}
observed in the cold atoms experiments \cite{garreau,garreau2}.
Thus the kicked rotator with one additional 
modulation frequency in time domain is similar to the case of 2D Anderson
localization in agreement with predictions done in 1983 \cite{dls1983}.
The case of 2 additional modulation frequencies 
is similar to the case of 3D Anderson transition
as discussed in \cite{borgonovi}.
We hope that the results presented here will
allow to investigate the Anderson localization
in 2D and 3D for periodically kicked rotator
with kicked incommensurate potential discussed in this work.

As it was pointed in Section I
the incommensurate standard map naturally appears for 
a description of dynamics of dark matter or comets in the Solar System
and other planetary systems with 2 or more planets
rotating around the central star. Recently
it has been shown that in the case
of star and one rotating planet the quantum effects can play a significant
role for escape of very light dark matter from the planetary system
due to the Anderson localization of energy transitions \cite{dlsdmp}.
The obtained results show that a presence of second planet 
leads to the dynamics described by the incommensurate
standard map with significant effects 
on the quantum localization of dark matter.

Since the Chirikov standard map has many universal features
and appears in the description of evolution of many
very different physical systems
we argue that the incommensurate standard map will also
find a broad field of applications.

%---------------------------------------------------------------------------
%%%%%%%%%%%%%%%%%%%%%%%%%
\section{Acknowledgments}
%%%%%%%%%%%%%%%%%%%%%%%%%
 This work was supported in 
part by the Programme Investissements
d'Avenir ANR-11-IDEX-0002-02, 
reference ANR-10-LABX-0037-NEXT (project THETRACOM).
This work was granted access to the HPC resources of 
CALMIP (Toulouse) under the allocation 2018-P0110.

%\end{acknowledgments}
%
%---------------------------------------------------------------------------
%%%%%%%%%%%%%%%%%%%%%%%

%
\end{document}